\documentclass[prl,reprint,superscriptaddress,longbibliography]{revtex4-1} %longbibliography,
\usepackage{epsfig}
\usepackage{amssymb}
\usepackage{threeparttable}
\usepackage{graphicx}
\usepackage{natbib}
\usepackage{longtable}
\usepackage{txfonts}
\usepackage{color}
\usepackage{lineno}

\begin{document}
%\linenumbers

\title{Sub-picosecond photo-induced displacive phase transition in two-dimensional MoTe$_2$}
\author{Bo Peng}
\affiliation{Key Laboratory of Micro and Nano Photonic Structures (MOE), Department of Optical Science and Engineering, Fudan University, Shanghai 200433, China}
\affiliation{Beijing National Laboratory for Condensed Matter Physics, and Institute of Physics, Chinese Academy of Sciences, Beijing 100190, China}
\affiliation{Cavendish Laboratory, University of Cambridge, J. J. Thomson Avenue, Cambridge CB3 0HE, United Kingdom}
\author{Hao Zhang} \email{zhangh@fudan.edu.cn}
\affiliation{Key Laboratory of Micro and Nano Photonic Structures (MOE), Department of Optical Science and Engineering, Fudan University, Shanghai 200433, China}
%E-mail: ,
%\$^{1\dag}$} 
\author{Weiwen Chen} 
\affiliation{Key Laboratory of Micro and Nano Photonic Structures (MOE), Department of Optical Science and Engineering, Fudan University, Shanghai 200433, China}
\author{Bowen Hou} 
\affiliation{Key Laboratory of Micro and Nano Photonic Structures (MOE), Department of Optical Science and Engineering, Fudan University, Shanghai 200433, China}
\author{Zhi-Jun Qiu}
\affiliation{State Key Laboratory of ASIC $\&$ System, School of Information Science and Technology, Fudan University, Shanghai 200433, China}
\author{Hezhu Shao} 
\affiliation{College of Electrical and Electronic Engineering, Wenzhou University, Wenzhou, 325035, China.}
\author{Heyuan Zhu}\email{hyzhu@fudan.edu.cn}
\affiliation{Key Laboratory of Micro and Nano Photonic Structures (MOE), Department of Optical Science and Engineering, Fudan University, Shanghai 200433, China}
\author{Bartomeu Monserrat}
\affiliation{Cavendish Laboratory, University of Cambridge, J. J. Thomson Avenue, Cambridge CB3 0HE, United Kingdom}
\affiliation{Department of Materials Science and Metallurgy, University of Cambridge, 27 Charles Babbage Road, Cambridge CB3 0FS, United Kingdom}
\author{Desheng Fu}\email{fu.tokusho@shizuoka.ac.jp}
\affiliation{Department of Electronics $\&$ Materials Sciences, Faculty of Engineering,  $\&$ Department of Optoelectronics and Nanostructure Science, Graduate School of Science and Technology, Shizuoka University, Hamamatsu, 432-8561, Japan}  
\author{Hongming Weng}% 
\affiliation{Beijing National Laboratory for Condensed Matter Physics, and Institute of Physics, Chinese Academy of Sciences, Beijing 100190, China}
\author{Costas M. Soukoulis}
\affiliation{Department of Physics and Astronomy and Ames Laboratory, Iowa State University, Ames, Iowa 50011, USA}

\begin{abstract}
Photo-induced phase transitions (PIPTs) provide an ultrafast, energy-efficient way for precisely manipulating the topological properties of transition-metal ditellurides, and can be used to stabilize a topological phase in an otherwise semiconducting material. Using first-principles calculations, we demonstrate that the PIPT in monolayer MoTe$_2$ from the semiconducting 2H phase to the topological 1T$'$ phase can be triggered purely by electronic excitations that soften multiple lattice vibrational modes. These softenings, driven by a Peierls-like mechanism within the conduction bands, lead to structural symmetry breaking within sub-picosecond timescales, which is shorter than the timescale of a thermally driven phase transition. The transition is predicted to be triggered by photons with energies over $1.96$\,eV, with an associated excited carrier density of $3.4\times10^{14}$\,cm$^{-2}$, which enables a controllable phase transformation by varying the laser wavelength. Our results provide insight into the underlying physics of the phase transition in 2D transition-metal ditellurides, and show an ultrafast phase transition mechanism for manipulation of the topological properties of 2D systems.  % paving the way for future 2D material-based science and technology.
% may trigger the broad interests in exploring the new physics  and applications of the low-dimensional systems.
\end{abstract}

\maketitle

A photo-induced phase transition (PIPT), resulting from cooperative electron-lattice interactions through transiently changing the electronic states of the solid by photoexcitations \cite{Nasu2004}, is completely different from thermally or pressure-induced phase transitions. A PIPT not only gives access to different %hidden 
phases in the solid to explore % the local non-equilibrium dynamics and
anomalous properties absent in the ground state phase \cite{Bisoyi2016}, but also enables precisely controllable phase transitions towards target structures with desired physical properties at high speeds \cite{Wu2011}. Thus, the initial discovery of a PIPT in organic charge transfer crystals has triggered great interest in a variety of fields \cite{Koshihara1990}. Compared with the vast number of materials exhibiting thermodynamic phase transitions, there are still very few solids undergoing a PIPT, and many investigations currently focus on bulk crystals possessing 1D correlated electron chains in which PIPTs may be easily triggered due to inherent instabilities caused by electron-electron and electron-lattice interactions in these quasi-1D systems \cite{Nasu2004,Hanamura1997}.

As an emerging platform for PIPTs, 2D and layered transition-metal ditellurides exhibit polymorphisms with distinct physical properties \cite{Wang2017d,Fei2018,Zhang2019}. Recent experimental evidence for a PIPT in MoTe$_2$ and WTe$_2$ shows that structural phase transitions often involve abrupt changes in the electronic structure such as the emergence of novel topological states \cite{Cho2015a,Sie2019,Yang2017a}. For instance, in few-layer MoTe$_2$, an irreversible transition from a semiconducting hexagonal 2H phase to a topological distorted octahedral 1T$'$ phase occurs under laser irradiation, which can be used to fabricate an ohmic heterophase homojunction with accurate control of micron-patterning in a desired area \cite{Cho2015a}. The phase transition is irreversible because the 1T$'$ phase formed in the electronically excited state is metastable, and to return to the ground-state 2H phase, the system has to overcome a potential barrier of around 0.71 eV \cite{Kolobov2016a,Krishnamoorthy2018}. The laser-induced 1T$'$ MoTe$_2$ is stable for more than one week in ambient conditions without any protection \cite{Tan2018}. Moreover, the structural distortion from the 2H to the 1T$'$ phase results in an intrinsic band inversion between Te $p$ and Mo $d$ bands \cite{Qian2014,Choe2016}, enabling ultrafast manipulation of the topological character of MoTe$_2$. The topological phase also displays gate-tunable superconductivity, providing a new potential platform to realize Majorana bound modes \cite{Sajadi2018,Fatemi2018}.

Despite all this progress, the microscopic nature of the photo-induced phase transition remains unclear. Triggered by external stimulation by light, the system undergoes changes in temperature, strain, electronic excitation, chemical state, and lattice vibrational modes. The transition may depend on one of these factors or on a combination of them \cite{Cho2015a,Kretschmer2017}. One proposal is that Te vacancies created by irradiation trigger the local phase transition \cite{Cho2015a,Wang2017c,Yoshimura2018}. Other theoretical and experimental studies have argued that accumulated heat is a main driving force for the phase transition \cite{Wang2017c,Tan2018}. Moreover, a strain-induced phase transition has been observed in monolayer MoTe$_2$ \cite{Duerloo2014,Song2016}, thus laser-induced thermal strain may also contribute to the observed phase transition. Yet another explanation is that it is the electronic excitation that plays a critical role in the phase transition \cite{Kolobov2016a,Krishnamoorthy2018}. Overall, the transition mechanism is still debated.

In this work, we demonstrate that the phase transition of monolayer MoTe$_2$ can be triggered by photoexcitation of carriers alone. Using first-principles calculations, we study the excited-state properties of 2H MoTe$_2$ at photoexcitation energies $E$ of 1.58 eV (785 nm), 1.96 eV (633 nm), 2.34 eV (532 nm), 2.42 eV (514 nm), and 2.63 eV (473 nm) by exciting all available electrons consistent with these energies. We find that a soft mode displacive phase transition occurs for $E>1.96$ eV. Photoexcitation modifies the electronic potential, resulting in a complete softening of the lattice vibrations that leads to spontaneous structural distortions with new equilibrium positions for the atoms. This structural transition takes place within one picosecond, shorter than the timescale for the photoexcited electrons to transfer their energy to the lattice. We also rule out a thermally-driven or a strain-induced phase transition under laser irradiation by comparing the thermodynamic stability of 2H and 1T$'$ MoTe$_2$ monolayers. Our findings not only reveal the potential origin of the PIPT in 2D transition-metal ditellurides, but also provide new insight into the observed dynamic transitions in their 3D counterparts, another question that remains unclear \cite{Sie2019,Zhang2019b,Kim2017}.% \textcolor{red}{BP: The electronic system is thermalized within $20$-$120$\,fs, while the phase transition has a timescale of $117$-$292$\,fs.} \textcolor{blue}{BM: is this your result or someone else's?} \textcolor{red}{BP: Yes, 20-120 fs comes from self-energy calculations, 117-292 fs comes from vibrational period.} We also rule out a thermally-driven or a strain-induced phase transition under laser irradiation by studying the thermodynamic stability of 2H and 1T$'$ MoTe$_2$ monolayers. Our findings not only reveal the origin of the PIPT in 2D transition-metal ditellurides, but also provide new insight into the observed dynamic transitions in their 3D counterparts, another question that remains unclear \cite{Sie2019,Zhang2019b,Kim2017}.

\subsection{Lattice mode softening} %Upon optical excitation, the electron distribution changes completely, and structural distortions hence arise.

\begin{figure}
\centering
\includegraphics[width=\linewidth]{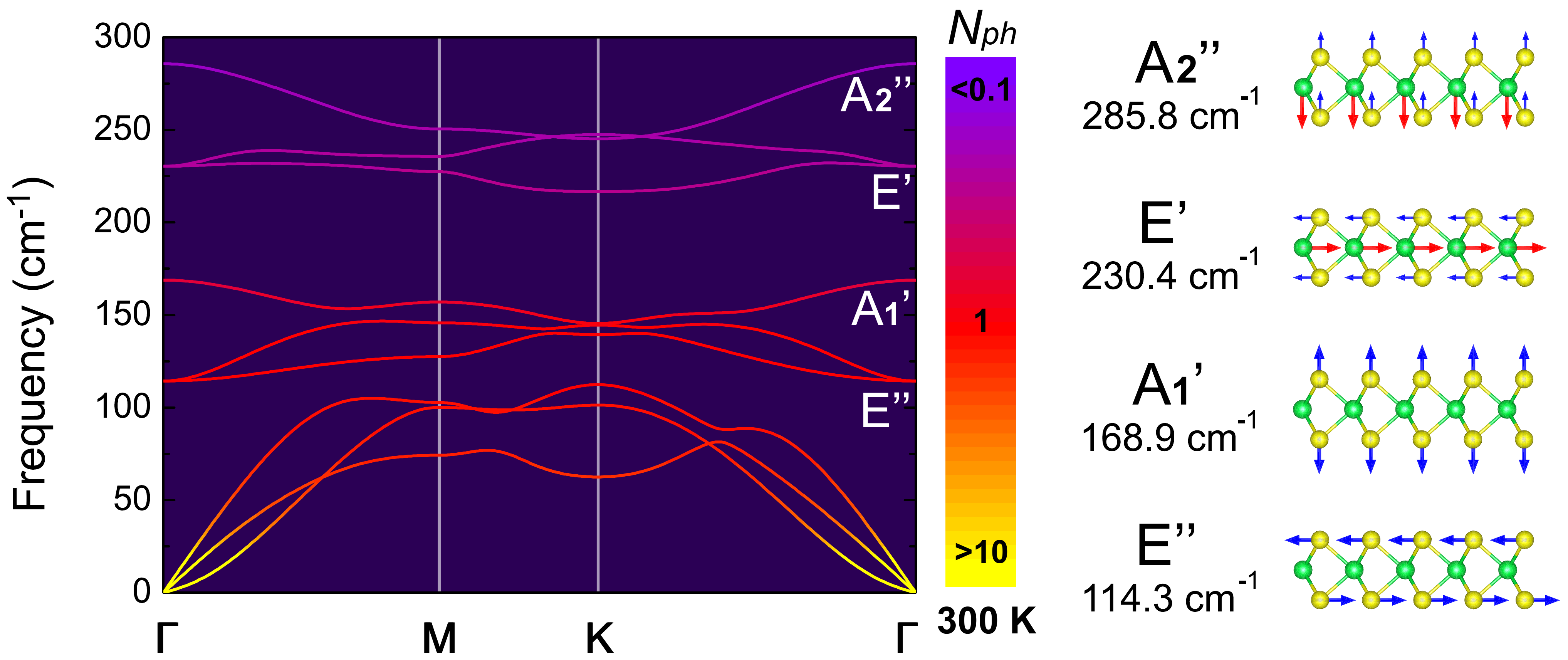}
\caption{Phonon dispersion of 2H MoTe$_2$ monolayer and vibrational modes for optical phonons at the $\Gamma$ point. The phonon occupation number is determined from the Bose-Einstein distribution function at 300 K.}
% (h) Schematic  potential energy wells of Landau theory for phase transition. (g) The change of parameter $a$ in Eq. (1) with excitation photon energy. }
\label{f2} 
\end{figure}

We start by examining atomic displacements in the excited state to see whether (and how) a phase transition may be triggered. With $D_{3h}$ point group, the irreducible representations of the vibrational modes in the 2H phase at the $\Gamma$ point are expressed as \cite{Yamamoto2014}
\begin{equation}
\Gamma_{\textrm{2H}} = \textrm{E}^{\prime\prime}\textrm{+A}_1^{\prime}\textrm{+E}^{\prime}\textrm{+A}_2^{\prime\prime}.
\label{eq1} 
\end{equation}
The calculated phonon dispersion is shown in Fig.~\ref{f2}, %with the symmetry labels of the optical phonon modes at the $\Gamma$ point marked
and the calculated lattice mode frequencies are consistent with previous reports \cite{Kan2015,Guo2015}.

\begin{figure}
\centering
\includegraphics[width=\linewidth]{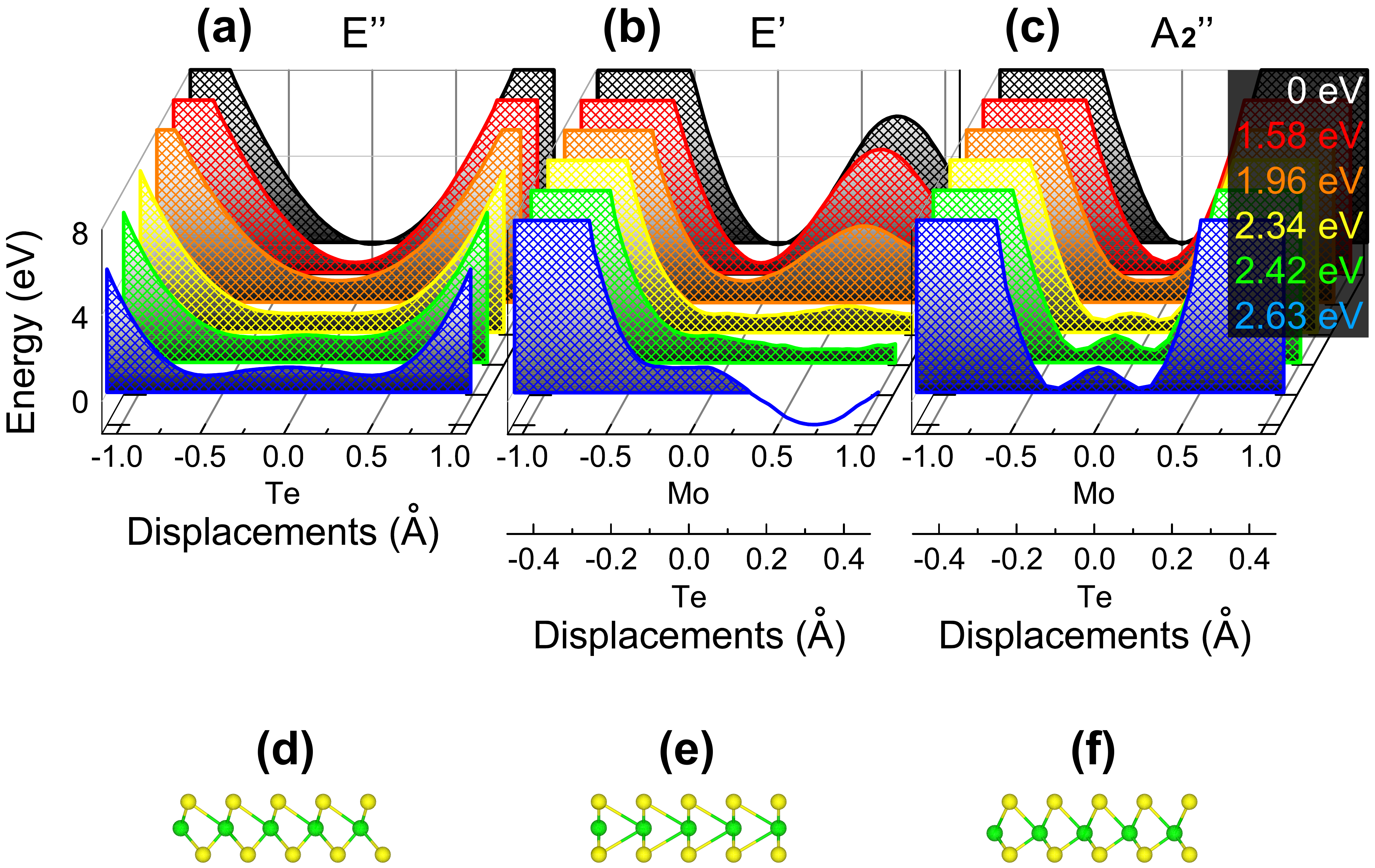}
\caption{Potential energy surface at different excitation energies along the eigenvectors of (a) E$''$, (b) E$'$, and (c) A$_2''$ modes, % The energies of the excited 1T$'$ crystal at different excitation energies are set as zero. 
and their corresponding structures (d)-(f) at the minima of the potential energy surface at 2.63 eV. % These intermediate crystal structures lead to the three primary lattice distortions along the pathway from the 2H to 1T$'$ phase.
}
% (h) Schematic  potential energy wells of Landau theory for phase transition. (g) The change of parameter $a$ in Eq. (1) with excitation photon energy. }
\label{f3} 
\end{figure}

The three modes E$^{\prime\prime}$, E$^{\prime}$ and A$_2^{\prime\prime}$ correspond to the three primary lattice distortions along the pathway from the 2H to the 1T$'$ phase, so could play a role in triggering the phase transition (for the three primary lattice distortions, see ``Crystal Structures" in {\sc{Supplementary Information}}). We calculate the potential energy as a function of displacement amplitude of these vibrations under different photoexcitation energies and corresponding carrier densities to investigate potential structural instabilities. The results are summarized in Fig.~\ref{f3}(a)-(c), with the energy of the excited 1T$'$ phase set to zero for comparing the relative stability between the 2H and 1T$'$ phases. 
In the ground state, the energy of the 2H phase is $80$\,meV lower than that of the 1T$'$ phase. However, as the system is optically excited, the 2H phase becomes energetically less favorable than the 1T$'$ phase due to population inversion \cite{Kolobov2016a,Krishnamoorthy2018}.

For the E$''$ mode (corresponding to the opposite in-plane motion of top and bottom Te layers), the Te atoms vibrate harmonically around their equilibrium positions before photoexcitation. With increasing excitation energy (and corresponding excited carrier density), the potential energy surface gradually flattens, and a double-well potential forms at $E>1.96$ eV. Thus, % the Te atoms have two energetically equivalent sites, and 
a distorted octahedral structure can form as atoms move to new equilibrium positions [Fig.~\ref{f3}(d)]. A similar situation occurs for the A$_2''$ mode (an out-of-plane displacement of Mo atoms and an opposite-direction displacement of Te atoms). For $E>1.96$ eV, the Mo atoms can bounce up or down to the two minimum energy sites of the new double-well potential, while the Te atoms move in opposite directions, leading to an out-of-plane distorted variant of 2H MoTe$_2$ [Fig.~\ref{f3}(f)]. The evolution of the E$'$ mode under irradiation is slightly different. This mode is composed of in-plane displacements with the single Mo layer and the two Te layers moving opposite to each other. Increasing $E$ lowers the energy barrier for positive displacements along the eigenvector of the E$'$ mode. The barrier becomes a deep well at 2.63 eV, leading to the atomic distortion shown in Fig.~\ref{f3}(e), with distorted in-plane atomic positions yielding a twisted octahedral coordination around Mo atoms. It should be noted that at 2.63 eV this mode dominates, but at lower energies the potential along the E$'$ mode still exhibits a barrier whereas both E$''$ and A$_2''$ modes show double-well potential energy surfaces, and therefore the latter two dominate in that regime. All three distortions of the corresponding mode eigenvectors are along the phase transition pathway from 2H to 1T$'$ MoTe$_2$. The full transition from 2H to 1T$'$ MoTe$_2$ probably occurs along a higher-dimensional path in configuration space in which photoexcitation lowers the energy barrier from 0.77 eV to 0.08 eV \cite{Krishnamoorthy2018}, but our results suggest that the softening of these key modes can initiate the PIPT with a purely electronic mechanism.

%The fact that these modes soften upon photoexcitation suggests a purely electronic mechanism for triggering the PIPT in MoTe$_2$. The full transition probably occurs along a more high-dimensional path in configuration space, but our results suggest but previous calculations have indicated that under photoexcitation, the energy barrier for the 2H to 1T$'$ transition drops from 0.77 eV to 0.08 eV \cite{Krishnamoorthy2018}.

% better packing of the Te-Mo-Te sub-lattice along the phase transition pathway from 2H to 1T$'$ MoTe$_2$, corresponding to the twisted octahedral coordination around Mo atoms.
%{\color{red} As previously reported, the 1T$'$ MoTe$_2$ phase becomes more stable under photoexcitation, , and these displacements tend to decay to the 1T$'$ phase. Previous calculations have indicated that under photoexcitation, the energy barrier for the 2H to 1T$'$ transition drops from 0.77 eV to 0.08 eV \cite{Krishnamoorthy2018}.}

\subsection{Critical excitation energy $E_C$}

\begin{figure}
\centering
\includegraphics[width=\linewidth]{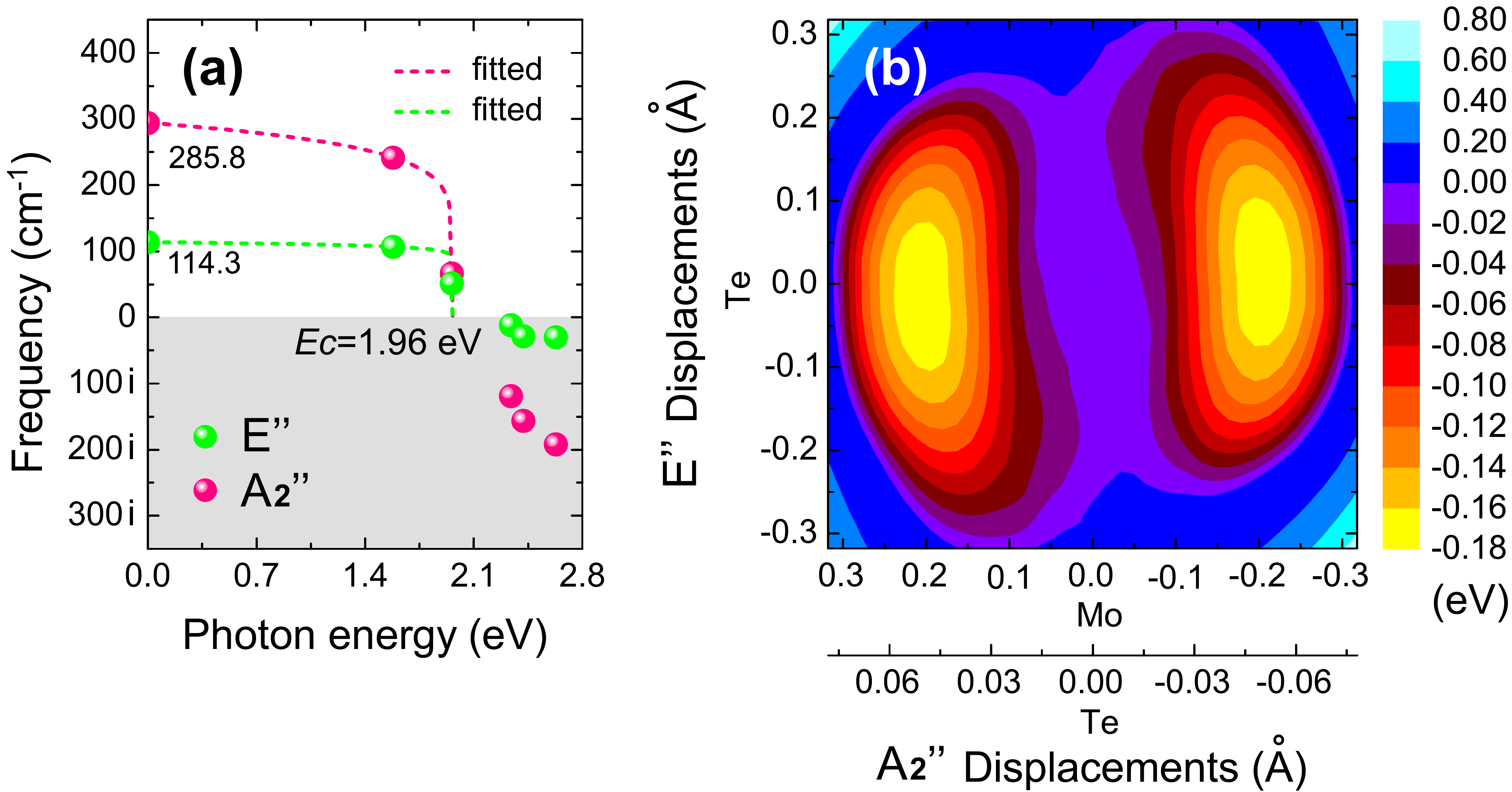}
\caption{(a) Phonon frequency of the E$''$ and A$_2''$ modes as a function of laser excitation energy. (b) Energy surface in the plane spanned by the two soft modes. Each contour line represents an energy increase of 0.02 eV.}
% (h) Schematic  potential energy wells of Landau theory for phase transition. (g) The change of parameter $a$ in Eq. (1) with excitation photon energy. }
\label{f4} 
\end{figure}

The transition from a single-well potential to a double-well potential shown in Fig.~\ref{f3}(a)-(c) indicates a displacive phase transition, which involves a phonon frequency that falls to zero at a critical excitation energy $E_C$ and corresponding excited carrier density. These zero-frequency soft modes drive the crystal instability along the displacements of the corresponding mode eigenvectors, leading to spontaneous symmetry breaking \cite{Dove1993,Dove2015}. We examine the soft-mode behaviour under optical excitation in the framework of Landau theory for the E$''$ and A$_2''$ modes that soften at lower energies. For a soft mode displacive phase transition, the order parameter can be chosen as the amplitude of the distortion of the soft mode eigenvector $u$ \cite{Dove1993}. The lattice potential energy, equivalent to the free energy in the Landau theory \cite{Kittel1976,Zeks1974}, can be expressed by a fourth order even polynomial expansion of the order parameter $u$
\begin{equation} \label{eq2} 
V(u) = a + \frac{b}{2}u^2+ \frac{c}{4}u^4,
\end{equation}
where the harmonic prefactor $b$ is equal to the square of the soft mode phonon frequency $\omega^2$ \cite{Dove1993,Monserrat2018}. Before the phase transition, there is a single-well potential, corresponding to $b>0$ ($c$ is a small positive value with a much weaker $E$ dependence). When $b$ becomes negative, $V(u)$ becomes a double-well function with minima at $u=\pm\sqrt{-b/c}$ and a local maximum value at $u=0$. Therefore, the phonon softening can be described quantitatively by calculating the phonon frequencies of the two symmetric E$''$ and A$_2''$ modes as a function of photoexcitation energy. As shown in Fig.~\ref{f4}(a), at 1.58 eV both modes soften slightly, and drop significantly at 1.96 eV. The phonon frequency becomes imaginary for $E>1.96$ eV, confirming again that a soft mode phase transition occurs at $E>1.96$ eV. It should be noted that at $E_C=1.96$ eV, the potential energy surface of the E$'$ mode is still a single well. The energy surface in the subspace spanned by both soft modes is calculated to further evaluate the spontaneous lattice distortions. Fig.~\ref{f4}(b) shows the energy surface mapped on the plane spanned by the two soft modes E$''$ and A$_2''$ at the photoexcitation energy of 2.34 eV. These results suggest that the A$_2''$ mode dominates the distortion, thus the PIPT is probably triggered along that distortion. 

%The phase transition occurs as soon as any one point on a phonon branch reaches zero. Similar to the Curie law in ferroelectric phase transition \cite{Kittel1976}, $\epsilon \propto (T-T_0)^ {\gamma}$, the critical excitation energy is fitted by $\omega \propto (E_C-E)^ {\gamma}$. For both the E$''$ and A$_2''$ modes, the transition excitation energy $E_C$ is the same (1.96 eV), indicating the crystal is unstable against the corresponding distortions. This is in good agreement with the fact that at excitation energies over 1.96 eV, the electron localization function shows distinct features such as shrinkage into the space between top and bottom Te atoms [Fig.~\ref{f3}(b)] and inverse triangle shape in the Mo layer [Fig.~\ref{f3}(c)]. Our prediction of photo-induced lattice mode softenings, marked by clear changes in both Raman and infrared spectra, can be easily verified experimentally by ultrafast coherent phonon spectroscopy \cite{Zhang2019b} or second harmonic generation \cite{Song2018a}.

\begin{figure}
\centering
\includegraphics[width=\linewidth]{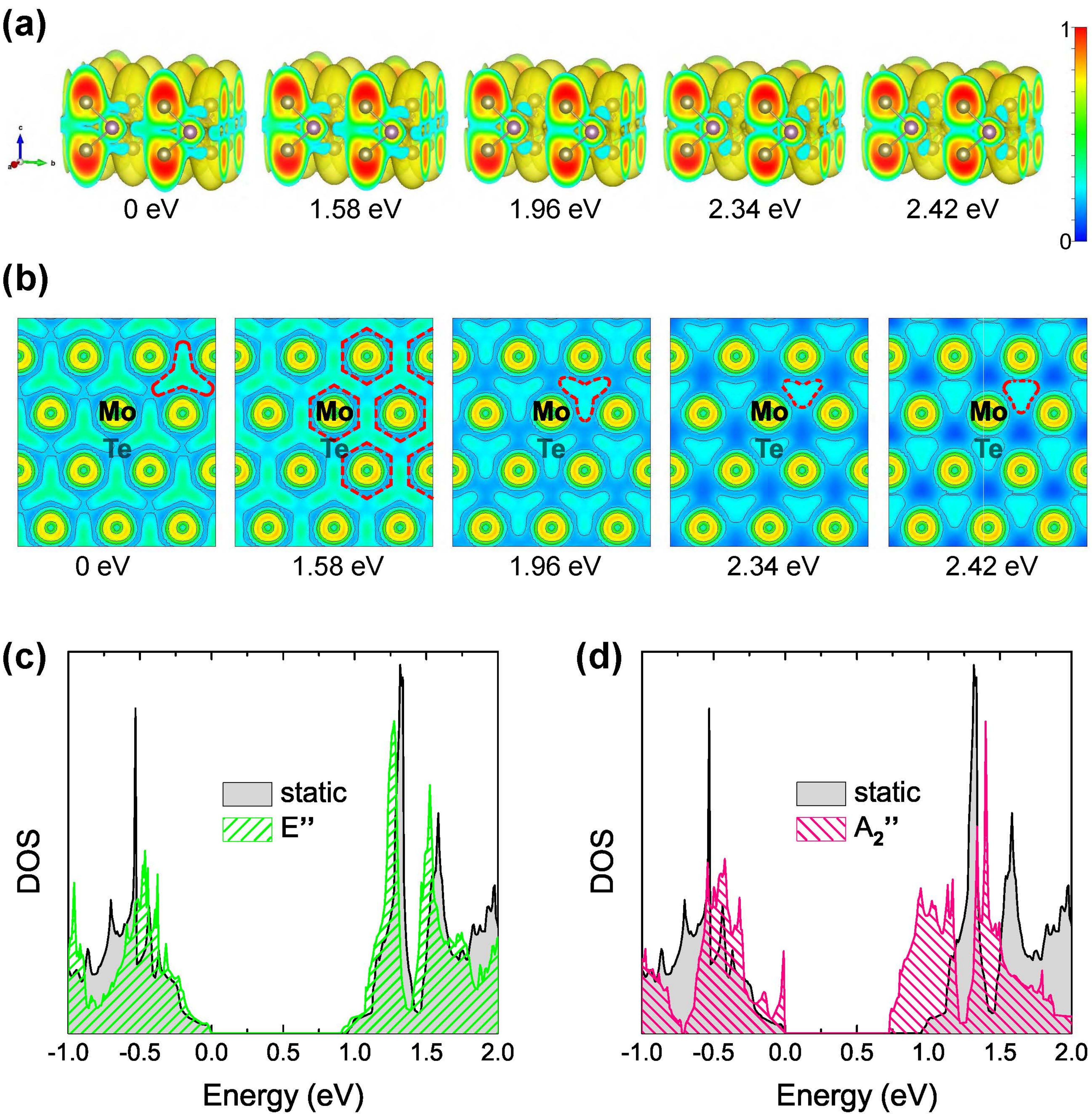}
\caption{(a) 3D and (b) top view (Mo layer)
% and (d) Te layer
of the electron localization function of 2H MoTe$_2$ at different excitation energies. The electron localization function from 0 to 1 is shown by colours from navy to red. The isosurface for the electron localization function in (a) is 0.2. Static lattice DOS (grey) compared to the DOS along the eigenvector of the (c) E$^{\prime\prime}$ (green) and (d) A$_2^{\prime\prime}$ (pink) phonon modes, with the displacement amplitudes corresponding to the potential minimum at 2.34 eV.}
\label{f1} 
\end{figure}

In the Landau theory, the phase transition occurs at $b=0$. For temperature-driven ferroelectric/magnetic phase transitions, $b \propto  (T-T_0)$, giving the well-known Curie-Weiss law \cite{Kittel1976}. In the detailed analysis of the potential well of the A$_2''$ mode, we find $b \propto (E_C-E)^ {\gamma}$, where $E_C=1.96$ eV and $\gamma=0.24$. Since $b=\omega^2$, we fit the mode frequencies as a function of $E$ by $\omega \propto (E_C-E)^ {\gamma/2}$, as shown in Fig.~\ref{f4}(a). The obtained $E_C=1.96$ eV is in good agreement with the change of electron distribution [Fig.~\ref{f1}(b)], from which one can see that the electron localization function also changes completely by $E\geq 1.96$ eV. A similar $E_C$ is also predicted for the E$''$ mode in Fig.~\ref{f4}(a). Thus our results suggest that the lattice distortions occur spontaneously after photoexcitation. Our prediction of photo-induced lattice mode softenings, marked by clear changes in both Raman and infrared spectra (E$''$ is Raman active and A$_2''$ is infrared active), can be verified experimentally by ultrafast coherent phonon spectroscopy \cite{Zhang2019b} or second harmonic generation \cite{Song2018a}.
% The predicted $E_c$ is also consistent  with the Raman scatter measurements, in which  coexistence of  2H and 1T' phase was observed  at photo-excitation of  $E=$1.96 eV while the   2H phase changes completely into  1T'  phase at $E>$1.96 eV.

% The difference between the fitting parameter $\gamma$ of thermally and photo-induced phase transition reflects their different mechanisms. 

For a thermally induced phase transition, $b=\tilde{b}(T-T_0)$, where $\tilde{b}$ is a positive constant \cite{Dove1993}, %and the high-order coefficient $c$ has a much weaker temperature dependence
the potential well changes gradually with temperature towards the phase transition. In contrast, in a PIPT, the potential well change is determined by the number of excited carriers, which is related to the excitation energy $E$ and the density of states (DOS). For instance, when $E$ is only slightly larger than the band gap, few electrons are excited to the conduction bands, leading to a negligible change of the potential well. Thus in the photo-induced phase transition, $b$ has a much weaker $E$ dependence than the $T$ dependence in a thermally induced phase transition, leading to a much smaller $\gamma$.

% \vspace{1mm}

% \noindent\textbf{Results and Discussion}

\subsection{Photo-induced changes in the electronic state}

We next investigate the electronic state to rationalize the phonon softening upon photoexcitation discussed above. When a solid undergoes a structural phase transition, the ionic positions change because the symmetry of the electronic potential changes \cite{Wall2012}. Light-induced electronic excitations in the 2H lattice may be responsible for such a change in the electronic potential.
 % \textcolor{green}{BP: This is describing a fact (electrons are excited under laser irradiation) and possible bond dissociation after irradiation.} %before the structural reorganizations, 
%and therefore we firstly examine how the photoexcitation changes the electronic state in the system. 
Fig.~\ref{f1}(a) shows the 3D electron localization function of 2H MoTe$_2$ at different photoexcitation energies. % calculated by promoting valence electrons to the conduction band to mimic optical excitation.
%{\color{red} With increasing excitation energy, the cyan contours are shrunk [as marked by red arrows in Fig.~\ref{f1}(b)], leading to distinct change of electron distribution in the Mo layer}. 
As shown in Fig.~\ref{f1}(b), the electron distribution around Mo atoms changes from isolated triangles (dashed red line) at 0 eV to connected hexagons (dashed red line) at 1.58 eV. As the excitation energy increases further to 1.96 eV, the electrons in the Mo layer gather between top and bottom Te atoms, and the in-plane triangles around Mo atoms appear once again, but with a reversed orientation compared to the ground state, indicating an electronic potential change. As the excitation energy increases to 2.34 eV, the inverted-triangle region of the electron density shrinks. 

%{\color{red} Thus the the degeneracy of electronic states may be further increased. According to the Jahn-Teller theorem, degenerate electronic states are unstable \cite{Jahn1937}, and such degeneracies can be removed by structural distortions. }

Figures~\ref{f1}(c) and (d) show a comparison of the DOS calculated for the static lattice and for the lattice distorted along the E$^{\prime\prime}$ and A$_2^{\prime\prime}$ phonon modes, with the displacement amplitudes corresponding to the potential minimum at 2.34 eV. As expected from the discussion in Fig.~\ref{f3} and~\ref{f4}, the A$_2^{\prime\prime}$ mode distortion leads to a larger change in the DOS: there is a significant redistribution of density of states in the conduction band towards lower energies. A similar picture emerges for the E$^{\prime\prime}$ mode, but in that case the changes are small. In the ground state the conduction band is empty and therefore a redistribution of the density of states towards lower energies does not affect the total energy of the system. However, upon photoexcitation, the conduction bands become occupied, and it then becomes energetically favourable to undergo a lattice distortion to red shift the conduction bands and lower the overall energy of the system. This is analogous to the Peierls mechanism, and provides a microscopic picture for the driving force behind the mode softening induced by photoexcitation. We note that the DOS around the valence band maximum also changes, and when those states are occupied in the ground state, the distortion leads to an energy increase as expected. However, under photoexcitation those states are unoccupied (holes), and the DOS redistribution has a minor effect on the total energy.

\begin{table}
\centering
\caption{Excited carrier concentration $n$ and the maximum amount of excitation $N$ by a laser pulse with a fluence of 100 mJ/cm$^2$, as well as their corresponding excitation percentage in the total number of valence electrons $N_{\textrm{tot}}$, at different laser excitation energies.}
\begin{tabular}{ccccccccc}
\hline
% & Space group & \multicolumn{2}{c}{Lattice constants} & \multicolumn{2}{c}{Bond length} & Cohesive energy & Band gap \\ 
 Energy (eV) & 1.58 & 1.96 & 2.34 & 2.42 & 2.63 \\
% Wavelength (nm) & 785 & 633 & 532 & 514 & 473 \\
\hline
 $n$ ($\times 10^{14}$ cm$^{-2}$) & 0.9 & 3.4 & 10.1 & 11.8 & 13.2 \\
 $n/N_{\textrm{tot}}$ (\%) & 0.8 & 2.9 & 8.5 & 9.9 & 11.1 \\
 $N$ ($\times 10^{14}$ cm$^{-2}$) & 7.2 & 12.2 & 16.0 & 17.2 & 17.6 \\
 $N/N_{\textrm{tot}}$ (\%) & 6.0 & 10.2 & 13.5 & 14.5 & 14.8 \\
\hline
\end{tabular}
\label{t1}
\end{table}

% \vspace{1mm}

\subsection{Excited carrier concentrations}

 % The PIPT is wavelength-dependent.
% The population of the occupied conduction bands at different excitation energies is determined by energy conservation, 
As discussed above, for a given photoexcitation energy we consider that the number of incoming photons is as large as necessary to excite all available electrons that can be excited at the given laser energy. Therefore, the excited carrier concentration becomes a function of the excitation energy. Table~\ref{t1} shows the excited carrier concentrations $n$ under different excitation energies. The excited carrier density at 1.96 eV %, under which the transition takes place, 
is $3.4\times10^{14}$ cm$^{-2}$, corresponding to an excited carrier density of 2.9\%. % A laser with an energy higher than 2.63 eV can readily generate excited carrier density in excess of 11.1\%. 
We also estimate the maximum amount of electron-hole pairs generated by a femtosecond laser with a fluence of 100 mJ/cm$^2$ assuming that no saturation occurs %\cite{Sokolowski-Tinten1995,Li2011}
[for the calculations of the maximum amount of electron-hole pairs, see ``Within 20 fs of Laser Excitation" in {\sc{Supplementary Information}}]. At 1.96 eV, excitation of up to 10.2\% is possible for a laser pulse with a fluence of 100 mJ/cm$^2$ (see Table~\ref{t1}), and thus the excited carrier concentrations in our calculations are experimentally accessible. As a PIPT can take place for excitations over 2.9\%, a laser pulse with a minimum fluence of 28 mJ/cm$^2$ is needed to induce a phase transition at 1.96 eV. This is consistent with the experimental fact that with insufficient laser fluence, the phase transformation cannot take place even for long irradiation times \cite{Cho2015a,Tan2018}.

%Our calculation also shows that phase change can take place for excitation over 9.9\% (2.42 eV). Thus a laser pulse with a fluence of 100 mJ/cm$^2$ is sufficient for the photo-induced phase transition. The minimum fluence for phase transition is calculated to be 68 mJ/cm$^2$.

%These laser energies correspond to high intensity excitations between the peaks in the DOS in Fig.~\ref{f2}(c).

\begin{figure}
\centering
\includegraphics[width=\linewidth]{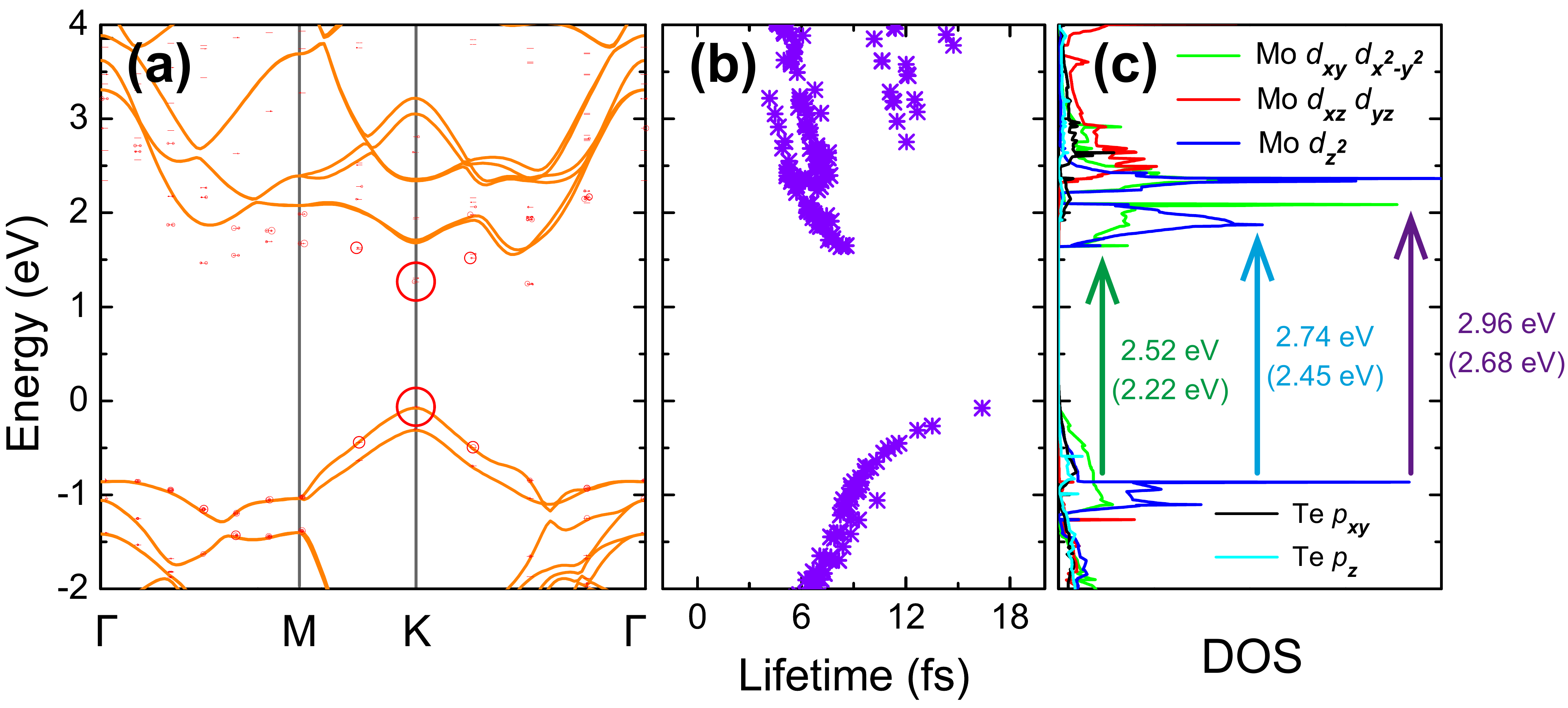}
\caption{(a) Electronic structures, (b) quasiparticle relaxation time, and (c) DOS of monolayer MoTe$_2$. The red circles in (a) denote the first exciton. The radius of circles represents the electron-hole coupling coefficient of excitonic wave functions. The larger the radius, the more an electron-hole pair contributes to the exciton eigenstate. The arrows in (c) indicate optical transitions with different electronic excitation energies. The parentheses in (c) show the exciton excitation energies. %(d) Imaginary part of dielectric function $\epsilon_2$, (e) dark, partially dark, and bright excitons with different binding energies, and (f) exciton lifetime for monolayer MoTe$_2$. The radius of circles in (e) represents the oscillator strength for bright excitons.
}
\label{f5} 
\end{figure}

The excited carrier population is related to the DOS \cite{Bernardi2014}. % {\color{red} The calculated $GW_0$ electronic structure is shown in Fig.~\ref{f5}(a). The calculated direct band gap at the K point of 1.76 eV is consistent with previous result of 1.77 eV \cite{Ramasubramaniam2012}. The calculated spin-orbital splitting for the valence band maximum at the K point is 239 meV, reproducing well the measured 250 meV \cite{Ruppert2014}.} As marked by arrows in Fig.~\ref{f5}(c), % the saddle points in the band structure are associated with large DOS. The electronic excitation energies of 2.52 eV, 2.74 eV, and 2.96 eV correspond to the optical transitions between the peaks in the DOS.
Optical transitions are dominated by peaks in the DOS, and the dominant peaks in MoTe$_2$ are marked by arrows in Fig.~\ref{f5}(c), where the optical transitions correspond to the excitation energies of 2.52 eV, 2.74 eV, and 2.96 eV.
In 2D materials, reduced electronic screening typically leads to strong excitonic effect. The Coulomb interaction of excited electrons and holes redistributes oscillator strength, giving rise to tightly bound excitons with eigenvalues lower than the band gap. By including electron-hole interactions, %the calculated imaginary part of dielectric function $\epsilon_2$ shows a non-zero value above the photon energy of 1.0 eV, which agrees well with the optical absorption measurement \cite{Ruppert2014,Lezama2015}. As shown in Fig.~\ref{f5}(d), strong optical absorptions at 1.39 eV and 1.89 eV reproduce well the measured large peaks. We plot the first 1000 excitons with their energy distribution. Dark excitons have zero oscillator strength, while bright excitons have large oscillator strength. The radius of circles in Fig.~\ref{f5}(e) represents the oscillator strength for bright excitons. The first exciton with a binding energy around 380 meV is bright. The red circles in Fig.~\ref{f5}(a) represent the electron-hole pairs that contribute to the first exciton. The larger the radius, the more contributions from the electron-hole pair. The eigenvalue of the first exciton, 1.37 eV, is larger than the measured optical band gap of 1.10-1.17 eV \cite{Ruppert2014,Lezama2015}. The measured optical transitions might originate from tightly bound trions \cite{Mak2012}, which cannot be described by the Bethe-Salpeter equation (BSE) \cite{Albrecht1998,Rohlfing1998,Sander2015}. Because of the electron-hole attraction,} 
the optical transition energies between the DOS peaks are reduced to 2.22 eV, 2.45 eV, and 2.68 eV [shown in parentheses in Fig.~\ref{f5}(c)]. With laser energies larger than 2.22 eV, the concentrations of excited carriers become higher, corresponding to strong optical excitation. % Moreover, the electronic excitation energies of 2.45 eV and 2.68 eV match well with the laser energies of 2.42 eV and 2.63 eV. Thus the excited carrier concentration $n$ at 2.34 eV is much larger that that at 1.96 eV.

\subsection{Timescale of the phase transition} 

The timescale after laser excitation is divided into three regimes: (1) excited electrons thermalization (in 20 fs), (2) soft mode phase transition (117-292 fs), and (3) lattice heating (within hundreds of ps).

In the first stage, the temperature of the electronic system increases by photoexcitation and the excited electrons thermalize within 20 fs at a high electronic temperature, while the lattice remains cold. In photoexcited MoTe$_2$, carrier-carrier scattering is responsible for redistributing carrier energy, resulting in a thermalized distribution function of carriers \cite{Shah2013}. The carrier relaxation time can be calculated from the quasiparticle self-energy \cite{Kotani2010,Peng2018c}. The excited electrons thermalize in 20 fs [Fig.~\ref{f5}(b)], whereas the electron-hole recombination occurs on the order of picoseconds [see ``Within 20 fs of Laser Excitation" in {\sc{Supplementary Information}}]]. Thus all the excited electrons are in thermal equilibrium before recombining or interacting with phonons. 
% The lifetime of these excited electrons, equivalent to the electron-hole recombination time, is calculated from the electron-hole amplitude and the quasiparticle self-energy \cite{Mak2014,Peng2018c}. As shown in Fig.~\ref{f4}(f), the recombination occurs on the order of picoseconds. Thus in the first stage (within 20 fs), all the excited electrons are in thermal equilibrium before interacting with phonons. 

In the second stage, the symmetry breaking displacements in 2H MoTe$_2$ take place within 292 fs. As the laser pulse creates electronic excitations, the new potential energy surfaces in Fig.~\ref{f3}(a)-(c) are formed immediately. The photo-induced lattice distortions, as a result of phonon softening, occur within the timescale of lattice vibrations. This can be estimated by the lattice vibration periods from 117 fs (A$_2''$ mode) to 292 fs (E$''$ mode). Such an estimate agrees well with previous molecular dynamics simulations \cite{Kolobov2016a} and pump-probe spectroscopy \cite{Zhang2019b}. The structural distortion can take place before the electrons transfer their energy to the lattice by carrier-phonon interaction (1-100 ps \cite{Shah2013}).

In the third stage, the hot carriers are cooled by transferring energy to the lattice, converting electronic photoexcitation energy to heat in hundreds of picoseconds through interaction with various phonons.

% \vspace{1mm}

\subsection{Competing mechanisms} 

One proposed mechanism for the PIPT in MoTe$_2$ is that after the excited electrons transfer their energy to the lattice through electron-phonon coupling, atomic reorganization toward the 1T$'$ phase may occur as a result of local heat \cite{Wang2017c} or thermal strain \cite{Duerloo2014,Song2016}. We next demonstrate that these thermal effects alone cannot trigger the phase transition.

\begin{figure}
\centering
\includegraphics[width=\linewidth]{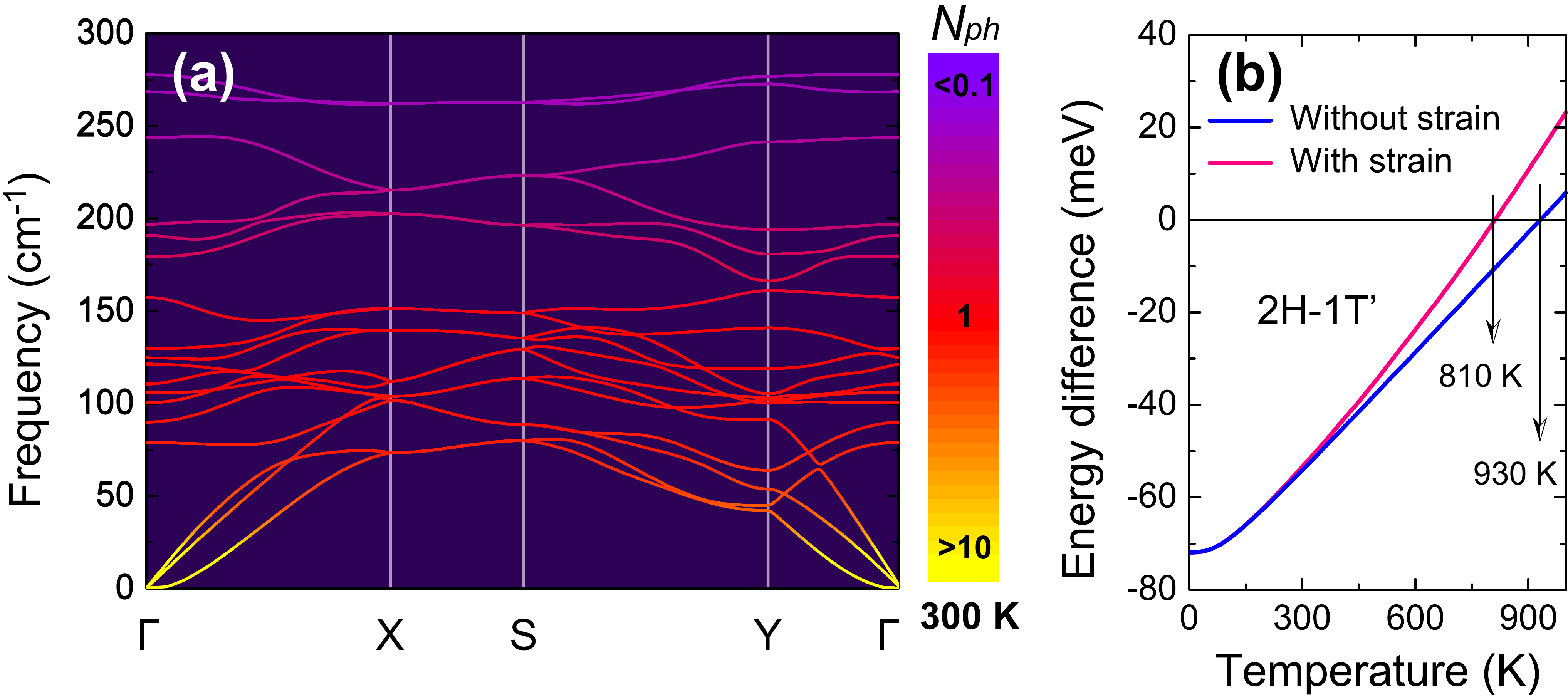}
\caption{(a) Phonon dispersion of 1T$'$ MoTe$_2$ monolayer. The phonon occupation number is determined from the Bose-Einstein distribution function at 300 K. (b) Helmholtz free energy difference between 2H and 1T$'$ MoTe$_2$ monolayers as a function of temperature with and without thermal strain.}
\label{f6} 
\end{figure}

As the thermalized electronic system transfers its energy to the lattice, local heat is generated. The estimated local temperature during the laser-irradiation is 670 K \cite{Cho2015a}. As the temperature increases, the phonons in the 1T$'$ MoTe$_2$ phase are more highly excited than the phonons in the 2H MoTe$_2$ phase, because the 1T$'$ phase has more low-energy phonons than the 2H phase [Fig.~\ref{f6}(a)]. Thus the entropy of the 1T$'$ phase becomes higher than that of the 2H phase as temperature increases (the entropy increases with the occupancy). It is thereby possible for the 2H structure to transform into 1T$'$ MoTe$_2$.

Thermodynamic stability at finite temperature can be described by the difference in Helmholtz free energy \cite{Setten2007,Peng2017a,Peng2018b} [for the calculations of thermodynamic stability, see ``Thermal Effects (about hundreds of ps)" in {\sc{Supplementary Information}}]. At 0 K, the free energy difference between the 2H phase and the 1T$'$ phase is just 72 meV per rectangular unit cell, indicating that the stability of the 1T$'$ phase is quite close to that of the 2H phase. As shown by the blue curve in Fig.~\ref{f6}(b), at temperatures higher than 930 K, 1T$'$ MoTe$_2$ becomes thermodynamically more stable than 2H MoTe$_2$. This is consistent with the fact that the pure 1T$'$ phase appears in a temperature range between 930 K (for Te deficiency) and 1170 K (for Te excess) \cite{Keum2015}. However, the temperature to facilitate the phase transition, 930 K, is significantly higher than the laser-induced temperature of 670 K \cite{Cho2015a}. Even after taking thermal strain into consideration [red curve in Fig.~\ref{f6}(b)], the phase transition temperature of 810 K is still higher than the laser-induced temperature [for the role of thermal strain, see ``Thermal Effects (about hundreds of ps)" in {\sc{Supplementary Information}}]. Therefore the laser-induced thermal effect on the phase transition has probably been overestimated in earlier studies \cite{Cho2015a,Wang2017c}.

% It should be noted that a transition from the 2H phase to the distorted phase can be triggered by purely electronic excitation when most energy remains within the excited electronic system. 

In fact, previous experiments have suggested that the heating effect alone cannot induce the structural transformation \cite{Tan2018}. For pure 2H MoTe$_2$, no significant change occurs after annealing; whereas laser irradiation leads to a mixed 2H + 1T$'$ phase. In the laser-irradiated mixed phase, the 2H phase disappears completely after annealing, leaving a pure 1T$'$ crystal \cite{Tan2018}. Therefore, electronic excitation may drive spontaneous lattice distortions, while the heating effect accelerates the phase transition via thermal displacements, in which the intermediate states decay to the 1T$'$ phase [for the role of thermal displacement, see ``Thermal Effects (about hundreds of ps)" in {\sc{Supplementary Information}}]. 

Another proposed mechanism for the PIPT in MoTe$_2$ is that laser irradiation can create Te vacancies, which can trigger a local phase transition \cite{Cho2015a,Wang2017c,Yoshimura2018}. Consistent with this picture, a recent study has demonstrated that the phase transition involves a set of successive dynamics such as Te vacancy diffusion and ordering to generate the 1T$'$ phase \cite{Si2019}, which is similar to the Ge-Sb-Te phase-change memory materials \cite{Zhang2012b}. A vacancy-induced phase transition is related to a change in chemical composition, which is analogous to the composition-dependent phase transition in Mo$_{1-x}$W$_x$Te$_2$ \cite{Lv2017a}. We emphasize that our work demonstrates that a phase transition can occur even in the absence of Te vacancies, and instead be driven by photoexcited carriers alone. Indeed, a recent experimental study has reported an electric-field induced structural transition in 2H MoTe$_2$ \cite{Zhang2019}, supporting our conclusion that changes in electron density alone can induce a phase transition without the need for Te vacancies. Similar photo-induced transitions have recently been predicted in the perovskites BaTiO$_3$ and PbTiO$_3$ \cite{Paillard2019}.

\subsection{Summary}

In summary, we demonstrate that a sub-picosecond phase transition in monolayer MoTe$_2$ from the 2H phase to a distorted 1T$'$ phase can be triggered by purely electronic excitation, even when the lattice is still cold. Such electronic excitation changes the electron density, resulting in a soft mode displacive phase transition. The phase transition from the 2H phase to the 1T$'$ phase can be induced by a critical excitation energy of 1.96 eV, at which 2.9\% of the carriers are excited. We also show that this phase transition mechanism can be understood in the frame of Landau theory, and is quite different from a temperature-driven phase transition. The microscopic picture of the phase transition is analogous to a Peierls distortion, with a red shift of the conduction bands that are occupied upon photoexcitation that lowers the overall energy of the system. Our findings shed new light on the microscopic origin of PIPT in 2D transition-metal ditellurides, and are expected to motivate both fundamental and applied studies of ultrafast phase transitions in these new class of materials for topological switching and neuromorphic computing.

% \vspace{1mm}

\subsection{Methods}

All calculations are performed based on density functional theory (DFT) using the Vienna \textit{ab-initio} simulation package (VASP) \cite{Kresse1996}. The exchange-correlation energy is calculated within the generalized gradient approximation (GGA) in the Perdew-Burke-Ernzerhof (PBE) parameterization \cite{Perdew1996}. In the excited state simulations, we mimick optical excitation by promoting valence electrons from high-lying valence band states to low-lying conduction band states. This method, typically called $\Delta$ self-consistent field ($\Delta$SCF) \cite{Jones1989,Goerling1996,Hellman2004}, introduces noninteracting electron-hole pairs by changing the occupation numbers of the Kohn-Sham orbitals and is less computationally demanding than more advanced approaches such as constrained DFT \cite{Mauri1995} and excited-state force calculations within $GW$+BSE \cite{Ismail-Beigi2003} but has been successfully used in studying PIPTs \cite{Li2011,Kolobov2016a,Chen2017a}. The occupied conduction band is determined by energy conservation at photoexcitation energies $E$ of 1.58 eV (785 nm), 1.96 eV (633 nm), 2.34 eV (532 nm), 2.42 eV (514 nm), and 2.63 eV (473 nm) (see ``Computational Details" in {\sc{Supplementary Information}}). We choose these laser excitation energies because the corresponding wavelengths are experimentally feasible. We keep these occupancies fixed throughout the calculations. This scenario corresponds to the time domain of interest, namely, after the laser-excited carriers populate the states near the band edges but before these carriers relax to the band edges via electron-phonon interactions. Electron-hole recombination occurs on the picoseconds timescale, which is much larger than a single phonon period and therefore the number of excited electrons can be reasonably assumed to remain constant as in the calculation. %Indeed, such approach has been used to study the excited state of monolayer and few-layer MoTe$_2$ \cite{Kolobov2016a}.
It should be noted that, for simplicity, we use the PBE band structures to determine the number of the excited electrons under optical excitation. The $GW_0$ band structure correction is an almost rigid upshift of the PBE conduction bands of 0.69 eV, and after a simple scissor shift, the required laser excitation energies are still experimentally feasible. %, because after taking excitonic effects the BSE absorption spectrum is similar to the PBE spectrum. 
In subsequent work, it would be interesting to see how bound excitations (excitons) modify the picture.

\vspace{6mm}

\noindent\textbf{Data Availability}

The raw data used in this study are available upon reasonable request from the corresponding authors.

\vspace{6mm}

%\noindent\textbf{References}

%, and Prof. Heejun Yang at Sungkyunkwan University

%\bibliography{new}




\vspace{6mm}

\noindent\textbf{Acknowledgements}

The authors gratefully acknowledge helpful discussions with Ms. Fangyuan Gu at Imperial College London and Prof. Xian-Bin Li at Jilin University. This work is supported by the National Natural Science Foundation of China under Grants No. 11374063, 11674369, 11404348 and 61774042,  the National Key Research and Development Program of China (No. 2016YFA0300600 and 2018YFA0305700), the ``Strategic Priority Research Program (B)" of the Chinese Academy of Sciences (Grant No. XDB07020100), and MEXT KAKENHI Grant Number JP17K05031, and Shanghai Municipal Natural Science Foundation (19ZR1402900, 17ZR1446500). 

\vspace{6mm}

\noindent\textbf{Author Contributions}

H. Zhang, Z-J. Qiu and H. Zhu designed the research. B. Peng and D. Fu performed the calculations. D. Fu, B. Peng, H. Zhu, W. Chen, B. Hou, H. Shao, B. Monserrat, H. Weng and C. M. Soukoulis analyzed and discussed the results. B. Peng, D. Fu and H. Zhang wrote the text of the manuscript.

\vspace{6mm}

\noindent\textbf{Competing Interests}

The authors declare no competing interests.

\vspace{6mm}

\noindent\textbf{Additional information}

Supplementary Information accompanies this paper at

[https://doi.org/]

Correspondence and requests for numerical data should be addressed to H. Zhang.

\end{document}